\documentclass[twocolumn, preprintnumbers, amsmath, amssymb, superscriptaddress, pra]{revtex4}
\usepackage{graphicx}
\usepackage{dcolumn}
\usepackage{bm}
\usepackage{epsf}
\usepackage{amssymb}
\DeclareGraphicsExtensions{.eps}
\usepackage{epstopdf}
\usepackage{graphicx}
\usepackage[sort&compress]{natbib}
\begin{document}

\author{David S. Simon}
\affiliation{Dept. of Physics and Astronomy, Stonehill College, 320 Washington Street, Easton, MA 02357}
\affiliation{Dept. of Electrical and Computer Engineering \& Photonics Center, Boston
University, 8 Saint Mary's St., Boston, MA 02215, USA}
\author{Nate Lawrence}
\affiliation{Dept. of Electrical and Computer Engineering \& Photonics Center, Boston
University, 8 Saint Mary's St., Boston, MA 02215, USA}
\author{Jacob Trevino}
\affiliation{Dept. of Electrical and Computer Engineering \& Photonics Center, Boston
University, 8 Saint Mary's St., Boston, MA 02215, USA}
\author{Luca Dal Negro}
\email[e-mail: ]{dalnegro@bu.edu}
\affiliation{Dept. of Electrical and Computer Engineering \& Photonics Center, Boston
University, 8 Saint Mary's St., Boston, MA 02215, USA}
\affiliation{Division of Materials Science \& Engineering, Boston University, Brookline, MA 02446, USA}
\author{Alexander V. Sergienko}
\email[e-mail: ]{alexserg@bu.edu}
\affiliation{Dept. of Electrical and Computer Engineering \& Photonics Center, Boston
University, 8 Saint Mary's St., Boston, MA 02215, USA}
\affiliation{Dept. of Physics, Boston University, 590 Commonwealth
Ave., Boston, MA 02215, USA}


\title{Quantum Key Distribution with Fibonacci Orbital Angular Momentum States}


\maketitle


{\bf Quantum cryptography and quantum key distribution (QKD) have been the most successful applications of quantum information processing, highlighting the unique capability of quantum mechanics, through the no-cloning theorem, to protect the security of shared encryption keys. Here we present a new and fundamentally different approach to high-capacity, high-efficiency QKD by exploiting interplay between cross-disciplinary ideas from quantum information and light scattering of aperiodic photonic media.
The novelty of the proposed approach relies on a unique type of entangled-photon source and a new physical mechanism for efficiently sharing keys. The new source produces entangled photons with orbital angular momenta (OAM) randomly distributed among Fibonacci numbers. Combining entanglement with the mathematical properties of Fibonacci sequences leads to a new QKD protocol. This Fibonacci protocol is immune to photon-number-splitting attacks and allows secure generation of long keys from few photons. Unlike other protocols, reference frame alignment and active modulation of production and detection bases are unnecessary, since security does not require use of non-orthogonal polarization measurements.}

Much recent work in QKD has shifted from use of two-dimensional polarization spaces to larger Hilbert spaces. Coding capacity and security increase with the size of the Hilbert space, and with the number of mutually unbiased bases used for security checks \cite{bruss,bechmann,bour,cerf,groblacher}. The most promising way to achieve larger Hilbert spaces is via optical OAM \cite{allen,mair}. However, the only practical way to produce entangled OAM states is with spontaneous parametric down conversion (SPDC), in which production amplitudes drop off rapidly with increasing OAM. In addition, eavesdropper detection in standard protocols requires active modulation of production and detection bases, slowing key generation rates exponentially as the range of OAM values increases.

Wavefront engineering of light traditionally relies on gradual phase shifts accumulated along optical beam paths. Recently, optical beams carrying single OAM states have been realized using planar plasmonic interfaces \cite{yu}. Additionally, distinctive scattering resonances carrying OAM have been demonstrated in nanoplasmonic Vogel spiral arrays \cite{trevino}, and the Vogel spiral geometry has been shown to support multifractal photonic band gaps with bandedge modes carrying multiple OAM values distributed among the Fibonacci numbers \cite{liew, trevino2}.
It has been analytically demonstrated that Vogel spiral arrays can generate multiple OAM states encoding well-defined numerical sequences in their far-field radiation patterns \cite{dalnegro}. In the case of golden angle (GA) spirals, the generated states carry OAM that follow the Fibonacci sequence. (Recall that the Fibonacci sequence \cite{koshy} obeys the recursion relation $F_n=F_{n-1}+F_{n-2}$, with initial values $F_1=1$ and $F_2=2$.)

Here, we combine GA spiral arrays with SPDC in a nonlinear crystal to
engineer a new type of entangled light source, producing photon pairs whose OAM values always sum to a Fibonacci number, allowing efficient production of states with large OAM values with properties that can be exploited in new ways.
We show that these special properties of entangled Fibonacci OAM states allow encryption keys with large numbers of digits to be generated by much smaller numbers of photons, exceeding the two bits per photon provided by quantum dense coding \cite{bennett}, while maintaining high security.
The approach uses entanglement, as in the Ekert protocol \cite{e91} (though it is used in a fundamentally different manner); however it requires one of the two legitimate users of the channel (Alice) to make her measurement before any opportunity for eavesdropping, as in the BB84 \cite{bb84} protocol.

{\bf Entangled Fibonacci spiral source.}
A Vogel spiral is an array of N particles with polar positions $(r_{n},\theta_{n})$ given in terms of scaling factor $a_{0}$ and divergence angle $\alpha$ by
\begin{equation}
r_{n} = \sqrt{n}a_{0}
\end{equation}
\begin{equation}
\theta_{n} = n\alpha
\end{equation}
An array of point scatterers, as in Figure 1(a), is then represented by a density function:
\begin{equation}
\rho(r,\theta) = \sum_{n=1}^{N} \delta(r-\sqrt{n}a_{0})\delta(\theta -n\alpha)
\end{equation}
We have previously shown that the Fraunhofer far-field of Vogel spirals can be calculated analytically, within scalar diffraction theory, for arbitrary $\alpha$ and $a_{0}$ \cite{dalnegro}. In cylindrical coordinates, the far-field of a diffracted input beam is given by \cite{dalnegro}
\begin{equation}
E_{\infty}(\nu_{r},\nu_{\theta}) = E_0\sum_{n=1}^{N}e^{j 2 \pi \sqrt{n} a_{0} \nu_{r} cos(\nu_{\theta}-n \alpha)}
\end{equation}
where $(\nu_{r}, \nu_{\theta})$ are the Fourier conjugate variables of $(r,\theta)$. As seen in Fig. 1(c), Fourier-Hankel analysis of the calculated far-field radiation pattern (shown in Figure 1(b)) is performed to decompose it into radial and azimuthal components, providing the OAM values \cite{liew,dalnegro,lawrence}. We see in Figure 1(d) that for GA spirals, OAM values are discretized in azimuthal numbers following the Fibonacci sequence. This follows directly from the geometrical properties of GA spirals encoded in the far-field patterns \cite{liew, dalnegro}. Fig. \ref{setupfig} then shows a schematic of our full QKD setup, in which the properties of the spiral source lead to a novel approach to high capacity QKD.

\begin{figure}
\centering
\includegraphics[width=3.5in]{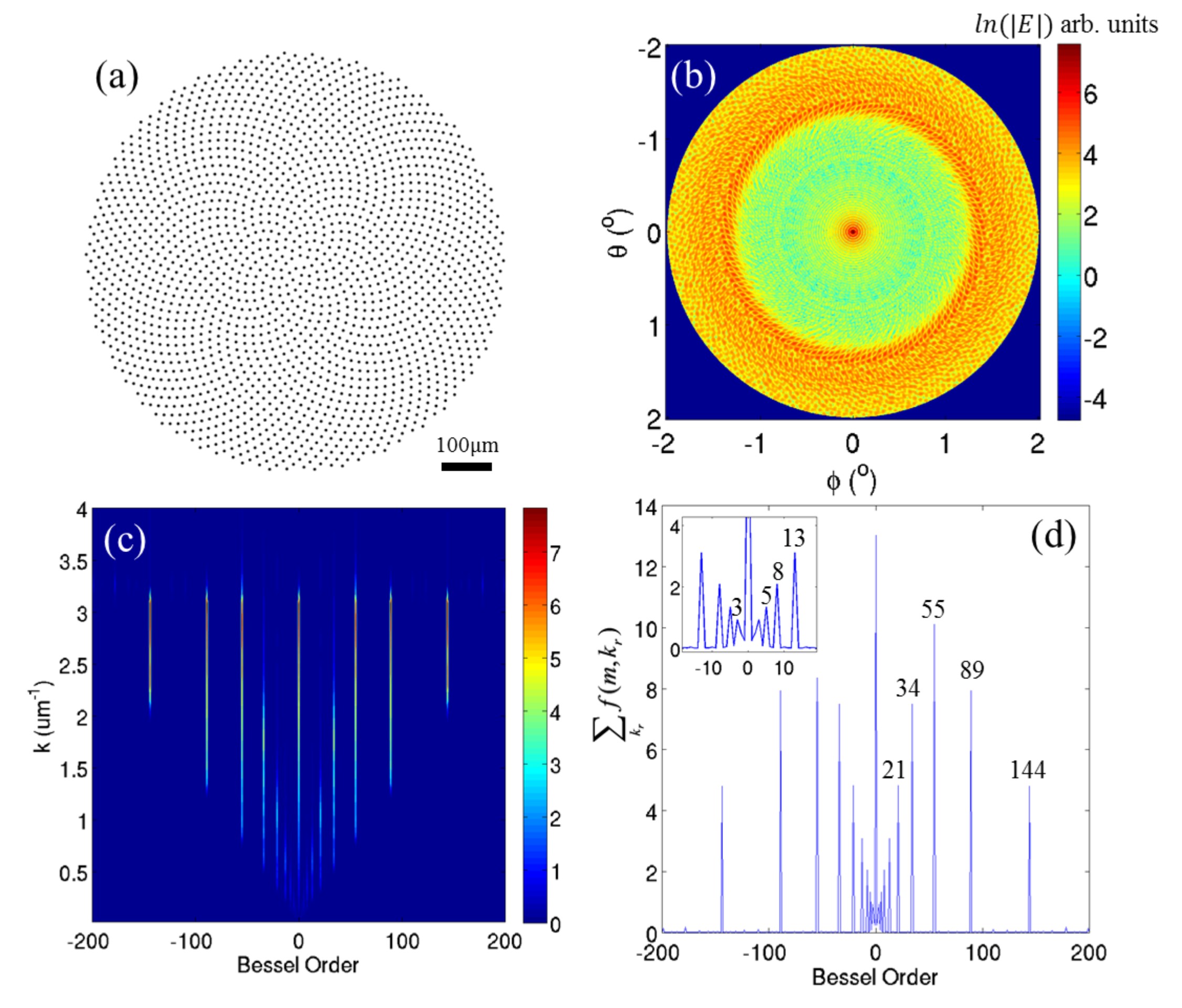}
\caption{{\bf (a)} Schematic of GA spiral Fibonacci OAM generator. {\bf (b)} Far-field pattern of GA spiral within a $2^\circ$ half-angle cone for a structure with 2000 particles and $a_0=9.28\mu m$ at $405\; nm$. {\bf (c)} Hankel transform of image in {\bf b}. {\bf (d)} Sum of {\bf c} over k, with peaks at Fibonacci values.}
\end{figure}

{\bf New QKD Protocol.}
In E91 \cite{e91} and BB84 \cite{bb84} protocols, photon polarization provides digits of a key (assigning, for example, $1$ to horizontal polarization and $0$ to vertical) and also provides security against eavesdropping: Alice and Bob each randomly pick one of two complementary bases in which to measure the photon polarization, keeping only photons for which the bases match. Eavesdropping is detectable by a drop in polarization correlations. OAM analogs of these protocols work in a similar manner, but with increased key generation capacity \cite{bechmann,bour,cerf,groblacher}, allowing multiple-digit segments of key to be transmitted by a single photon.

The light coming from the spiral will be in a superposition of states with OAM equal to Fibonacci numbers.
For the new protocol, we choose $N$ consecutive values, ${\cal F}=\left\{ F_{n_0},F_{n_0+1},\dots ,F_{n_0+N-1}\right\} $, and assign a block of binary digits to each so that equal numbers of 0's and 1's occur. If OAM values in this set are used, each photon generates enough digits to encode $\log_2 N$ bits of information. Here, we assume $N=8$ to illustrate the potential for high capacity. For example, the Fibonacci numbers from 3 to 89 may be assigned three-digit blocks as follows:
\begin{equation}\begin{array}{ccccccc}
3= 000 & \qquad &   8= 010 &  \qquad  & 21= 100   &\quad  & 55=  110\\
5= 001& &          13= 011 &         &       34=  101 & & 89= 111\\
\end{array}\label{binarylist}\end{equation} Three key digits are then carried by the OAM of a single photon.  The values in ${\cal F}$ must be arranged to be detected with equal probability; see appendix A.
The SPDC spiral bandwidth (range of OAM values) must be sufficient to span the largest gap in ${\cal F}$. Bandwidths over $40$ have been achieved \cite{romero}, so the values used here are currently practical. Greater bandwidths allow larger sets ${\cal F}$, increasing both information capacity and security. For simplicity, we assume that OAM sorters \cite{leach,berkhout,lavery} only allow positive OAM values to reach the detectors. (Including negative values doubles the capacity for the same ${\cal F}$; see appendix D.)

\begin{figure}
\centering
\includegraphics[totalheight=2.0in]{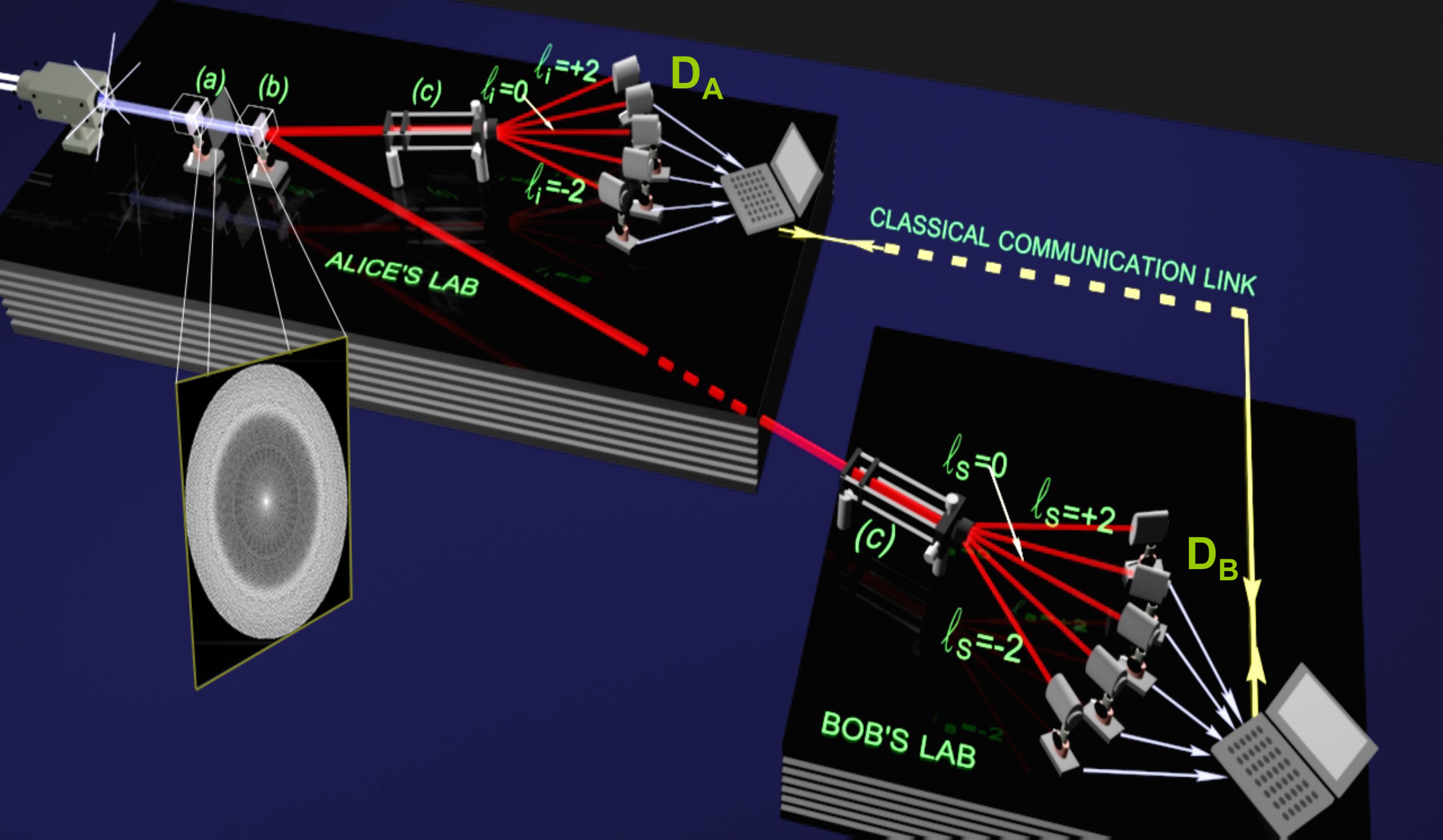}
\caption{Setup for QKD with Fibonacci-valued OAM states.
A laser interacts with a spiral array (a) , producing intense Fibonacci OAM states with $l=F_n$ that are used to pump the nonlinear crystal (b), producing signal-idler pairs through SPDC.
The OAM sorters (c) are arranged to only allow photons to reach the arrays ($D_A$ and $D_B$) of single-photon detectors if they also are Fibonacci-valued, with OAM $F_{n_i}$ and $F_{n_s}$. These must add up to the pump value: $F_n=F_{n_i}+F_{n_s}$. Only pairs of values $F_{n_i}$ and $F_{n_s}$ which are between $1$ and $54$ and which sum to $F_n$ values between $3$ and $89$ are kept.}\label{setupfig}
\end{figure}

\begin{figure}
\centering
\includegraphics[totalheight=2.2in]{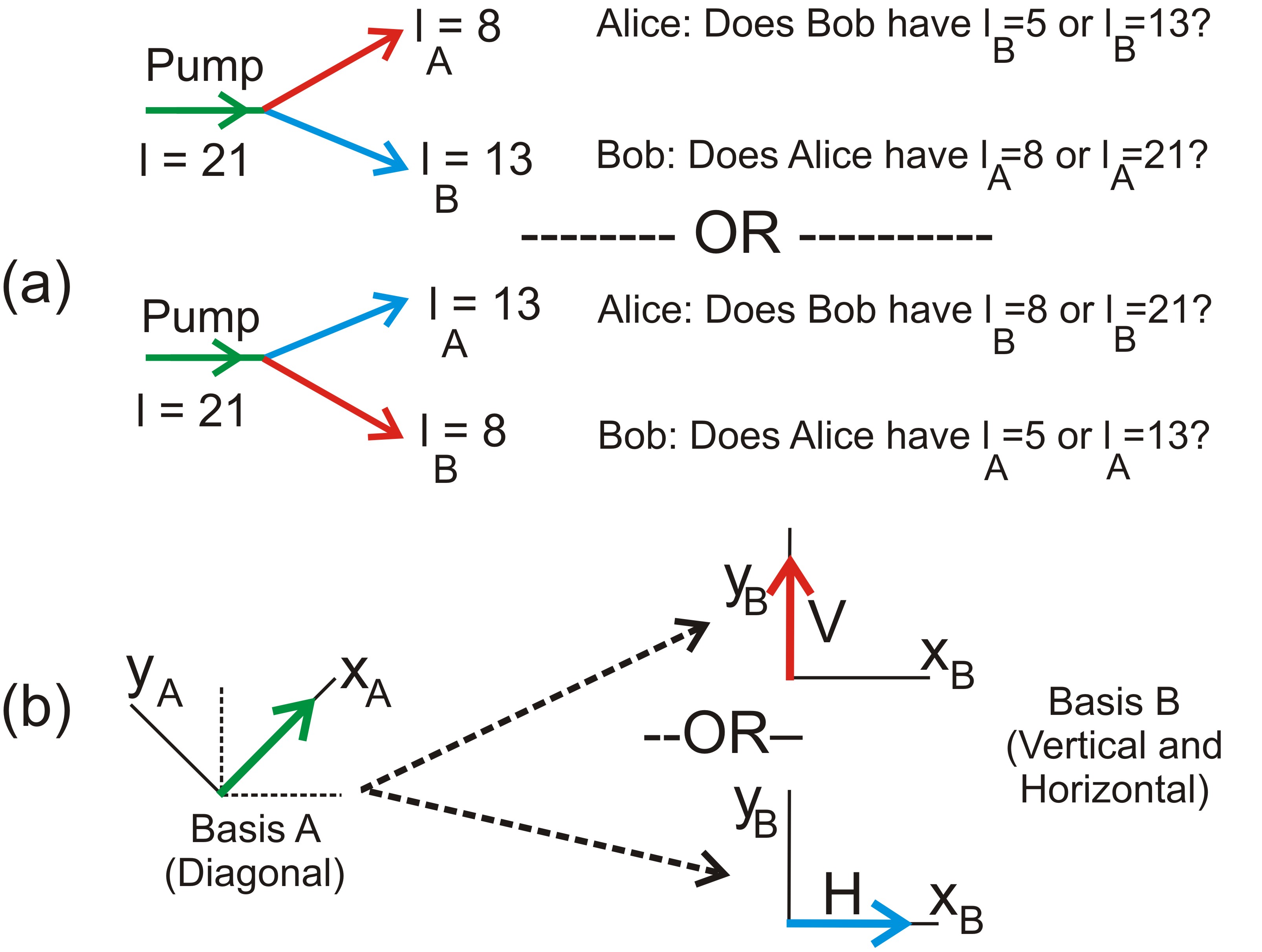}
\caption{Possible outcomes {\bf (a)} for the case where the total OAM is $l=21$. Neither Alice nor Bob knows the value received by the other; each knows that the two transmitted values must be adjacent Fibonacci numbers, but neither knows if the other's value is larger or smaller than their own. This replaces the ambiguity introduced in standard protocols by the nonorthogonality of the possible polarization bases {\bf (b)}, where a vector along one axis in the $A$ basis could be measured along {\it either} axis in the $B$ basis. }\label{protocolfig}
\end{figure}

\begin{figure}
\centering
\includegraphics[totalheight=2.0in]{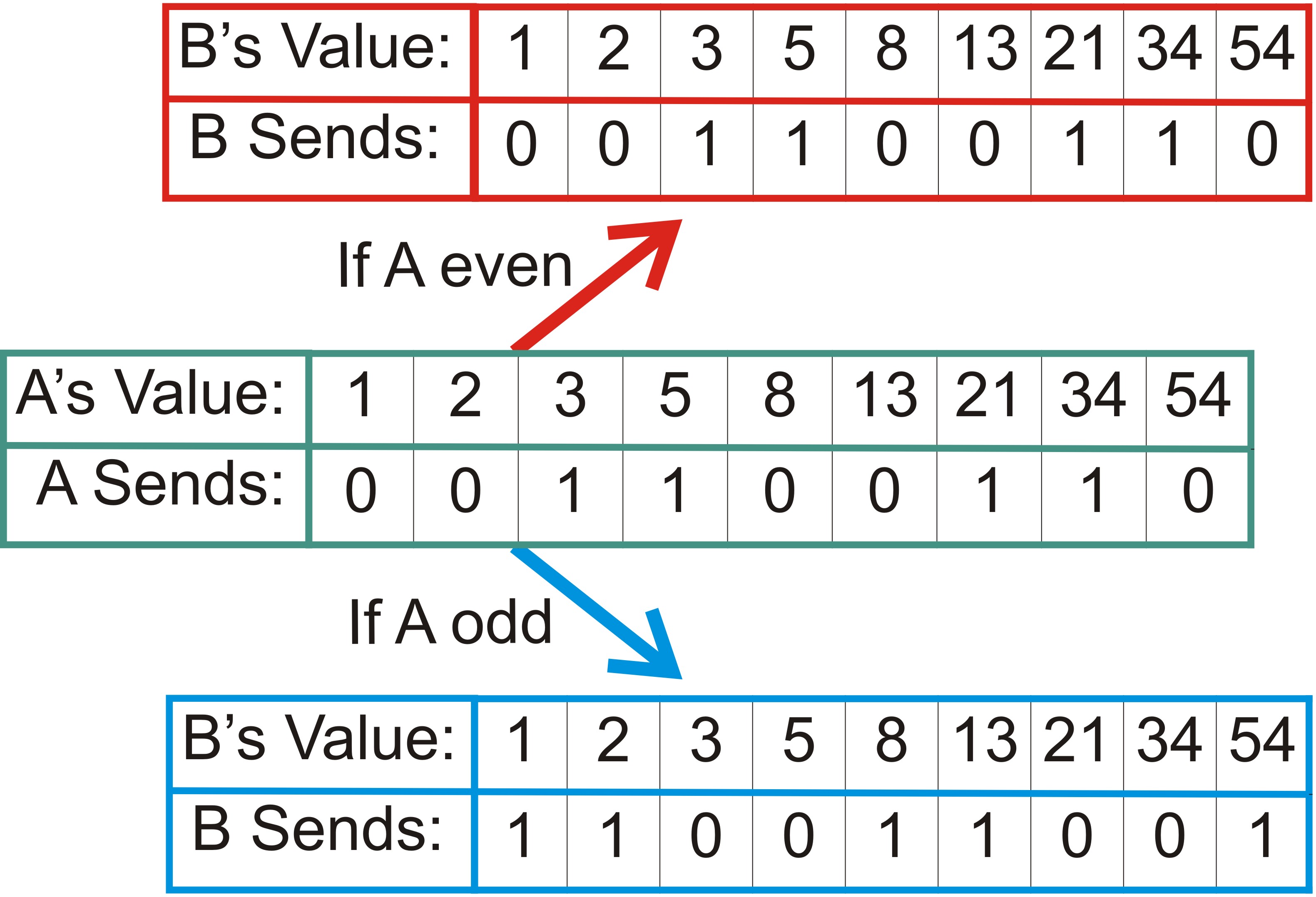}
\caption{Scheme for classical information exchange. Alice first sends Bob one bit according to the scheme in the middle row. Since Bob knows Alice must have either the Fibonacci number before his or the one after, this is sufficient for him to deduce her value. Bob then transmits one bit to Alice. To minimize Eve's ability to reconstruct the values, Bob uses the same scheme Alice did when her value is even (top row), but uses the conjugate scheme ($0$'s and $1$'s interchanged) when her value is odd (bottom row). Both now know each other's values, and can add them to get the value $l=F_n$ of the pump, but an eavesdropper cannot (see appendix B for more detail). $F_n$ then serves as the key segment.}\label{tablefig}
\end{figure}

Imagine a photon with OAM in ${\cal F}$ (take $l=F_n=21$ as an example) entering the crystal. The resulting signal and idler OAM, $l_i$ and $l_s$, are not necessarily Fibonacci, but the OAM sorters may be arranged to allow
only photons with OAM in ${\cal F}$ ($l_i=F_{n_i}$, $l_s=F_{n_s}$) to register.
For collinear SPDC (type I or type II) OAM conservation implies
$F_{n_i}+F_{n_s}=F_n$. The Fibonacci recursion relation forces $F_{n_i}$ and $F_{n_s}$ to be the two Fibonacci numbers immediately preceding $F_n$ ($F_{n-2}=8$ and $F_{n-1}=13$ in our example).  However, which reaches Bob and which reaches Alice is undetermined, so there are two possibilities (Figure \ref{protocolfig} (a)). Suppose Bob receives $l_i=8$ and Alice receives $l_s=13$. Then Alice doesn't know if Bob has $8$ or $21$ (the Fibonacci value before hers, or the one after). Similarly, Bob doesn't know if Alice has $5$ or $13$. To determine each other's values, each must send one classical (potentially public) bit to the other (see Fig. \ref{tablefig}). They then add their values to get the pump value $F_n=21$, which serves as one segment of the key.

The protocol utilizes two complementary sources of ambiguity for secure communication: uncertainty in how the OAM Fibonacci state is decomposed between Alice and Bob minimizes the amount of information an eavesdropper could obtain from the classical exchange (see appendix B), whereas Eve reveals her presence through her failure to uniquely identify a particular OAM value in the quantum channel due to uncertainty as to which of two possible superpositions it originates from (see below). Detailed security analysis will be conducted elsewhere, but appendix C points out the inherent immunity to photon-number-splitting attacks and the possibility of a new type of decoy state.

{\bf Detecting eavesdropping.}
The usual method of detecting eavesdroppers employs two (or more \cite{bruss,bechmann,bour}) mutually unbiased bases (Figure \ref{protocolfig} (b)). However, in the current setting a more novel possibility arises. Photons leaving the spiral are in superposition states: $|\Psi_0\rangle ={1\over \sqrt{N}} \sum_n |F_n\rangle .$  Down conversion splits each $|F_n\rangle$ into a state of form
$|\Psi_1\rangle =C_1\sum_{n,l}|l\rangle_A |F_n-l\rangle_B .$ The sorters pick out values $l\in {\cal F}$, leading to entangled combinations of consecutive Fibonacci numbers: $|\Psi\rangle =C_2\sum_n \left\{|F_{n-1}\rangle_A |F_{n-2}\rangle_B +|F_{n-2}\rangle_A|F_{n-1}\rangle_B\right\} .$ Assume that
the entangled photons are created in lab $A$ and Alice measures her value immediately, while the other photon is still in transit to lab $B$.

If Alice measures $l_A=8$, the state reaching Bob will be a superposition, ${1\over \sqrt{2}}\left\{ |5\rangle_B +|13\rangle_B\right\} .$
Suppose Eve intercepts Bob's photon and measures $l_B=5$. She then resends a new photon to Bob
in place of the one she intercepted. She knows she should transmit a superposition, but she must guess {\it which} superposition to prepare: ${1\over \sqrt{2}}\left\{ |5\rangle_B +|13\rangle_B\right\} $ or ${1\over \sqrt{2}}\left\{ |1\rangle_B +|5\rangle_B\right\} $? So when Bob makes his measurement, he will find the possible values $1$, $5$, or $13$, with respective probabilities ${1\over 4}$, ${1\over 2}$, and ${1\over 4}$. But, given Alice's measurement of $8$, the only values he {\it should} be receiving are $5$ and $13$, with probability $1\over 2$ each. So the security protocol has Alice and Bob telling each other their values for a randomly selected subset: if Eve has been at work, $25\% $ of the time they will have values which are not adjacent (like $1$ and $8$ in the example above).
If no problem appears in the security subset, the remaining photons then generate the high capacity key:
if $n$ photon pairs are left after security checks, Alice and Bob share a $3n$-digit key. Generalizing from three-digit segments per photon to longer segments involving larger $F_n$, is obvious.

{\bf Advantages.}
Besides increased capacity per photon, the Fibonacci protocol has several advantages: (i) Intercepting one of the photons by itself does not allow Eve to determine that key segment, since each photon only carries half the information needed to reconstruct the key.
(ii) Carried out in free space, irrelevant photons coming from the ambient light tend to be automatically screened out, since only photons with Fibonacci-valued OAM contribute. (iii) Randomized OAM values are produced in a completely passive manner, without need for active switching of holograms, as required by other OAM-based QKD approaches, greatly speeding up key generation rates. (iv) The Fibonacci numbers have gaps between them, reducing misattribution errors. (v) Fibonacci coding can be more efficient than binary coding for some purposes \cite{klein}. (vi) Unlike in polarization-based QKD, no alignment of reference frames is needed. (vii) The procedure is, in principle, highly scalable: to use at longer distances, the crystal may be pumped at higher intensity and the alphabet of Fibonacci numbers ${\cal F}$ can be increased in size, allowing the same rate of key generation despite increased losses. The use of multiple parallel detectors reduces speed limitations on the detection side.

One disadvantage is that the classical exchange includes information about the actual key (not just about measurement bases, as in other protocols). If Eve intercepts it, she can narrow possibilities for each three-digit segment from $8$ down to three or four (appendix B). She cannot determine the value uniquely, but preventing Eve from obtaining {\it any} information about the key at all is highly desirable. A variation of the protocol which eliminates this weakness is in fact possible, at the cost of increased complexity; a description of this is in preparation.

Finally, the protocol depends only on the recurrence relation, not on the starting values of the sequence. So an identical protocol exists, for example, for the Lucas sequence \cite{koshy}. More generally, other two-term recursion relations may also be used.

We have demonstrated a new realization of high-capacity, high-efficiency quantum cryptography, based on specially engineered OAM-entangled states of light and a new QKD protocol exploiting recursive properties of the Fibonacci sequence. We believe this approach is general enough to lead to novel QKD implementations using other physical variables, such as encoding Fibonacci numbers in phase.

\appendix

\renewcommand*{\thesubsection}{A\arabic{section}}
\renewcommand*{\thefigure}{A\arabic{figure}}
\renewcommand*{\theequation}{A\arabic{equation}}

\section{Equalization of probabilities}If some of the values in the chosen set of Fibonacci numbers are more likely than others to contribute to the key, this bias provides a loophole that the eavesdropper can use to compromise security. Since neither the golden angle spiral nor the down conversion process have flat distributions in the space of angular momentum values, we must somehow equalize the probabilities of the $F_n$ values detected. This can be done in several ways. First, by appropriately engineering the spiral source and altering the distance of the spiral from the down conversion crystal, there is a measure of control over the spectrum that in principle can be used to flatten the OAM spectrum. However, a simpler solution is simply to insert filters in the apparatus. Filters can be designed that allow some angular momenta to pass with higher probability than others, and so these can be used to compensate for the distribution in the signal and idler beam after the crystal. A still simpler solution is to place filters before the detectors. The sorters convert different angular momentum values into different spatial locations, so a filter of appropriate transmission probability can be placed at each position. This last possibility has at least two additional advantages: (i) Each filter need only be chosen to transmit a particular proportion of the total intensity; its transmission profile does not need to be OAM-dependent. (ii) This adds an additional contribution to security, since even when Eve intercepts a photon and resends a copy, she does not know if the one she sends will be one of those that survive the filtering process at the end, thus essentially adding noise to Eve's signal, but not to Alice's or Bob's. This reduces her effective eavesdropping rate.

\section{Security from eavesdropping}
In the main text, a scheme is described in which Alice and Bob exchange one bit each of classical information in order to determine each other's values. They then add the two values together to get the orbital angular momentum $l=F_n$ of the pump, and that total value serves as a segment of the key. Here we point out that for an eavesdropper listening in on the classical channel, the information exchanged is insufficient to determine the value. Using the procedure that was outlined, each classical exchange leads to ambiguous results for Eve:
\begin{equation}\begin{array}{r|c|c|c|c} \mbox{Eve sees:} & 00 & 01 & 10 & 11\\ \hline
l\mbox{ could be:} & 3, 21, 34, 89 & 3, 5, 13, 21 & 8, 55, 89 & 5, 13, 34, 55 . \end{array}\end{equation} (In the top row, the first digit in each pair is the classical bit sent by Alice, the second is the bit sent by Bob.)
Note that each $l$ value except $8$ can be represented by two different classical exchanges, and that each exchange can represent three or four different $l$ values: if Eve intercepts the classical exchange, she has  a probability of only $1\over 4$ to $1\over 3$ of correctly guessing the value of $F_n$, with the {\it average} probability of a correct guess being $27.08\%$.  The probability of a correct guess drops as the number $N$ of Fibonacci values used increases. Alice and Bob can determine each other's values, while Eve cannot. This is possible only because of the combined action of (i) the entanglement and (ii) the properties of the Fibonacci recursion relation.

Because of the different nature of the ambiguity in the Fibonacci QKD protocol, a strong-light analog of the proposed approach could be devised. A computer could be programmed to randomly change settings on a pair of spatial light modulators or other device in order to encode pairs of adjacent Fibonacci values on the OAM content of two strong Gaussian pulses (with all the photons in each pulse being in the same OAM state). The ambiguity in who has the higher of the two values in the Fibonacci number decomposition allows the random key to be shared in the same manner as before. Eve would still not be able to obtain significant information about the key from the public exchange between Alice and Bob. However, if Eve now intercepts the OAM state sent to Bob (on what was previously the quantum, single-photon channel), the eavesdropping can no longer be revealed, since Eve may now siphon off multiple photons from the beam to reconstruct its state exactly without changing the state of the remaining beam. In other words, secure key generation with minimal information leakage through the public exchange can be performed with Gaussian states by exploiting Fibonacci recurrence, but the confident detection of an in-line eavesdropper's presence would require the quantum regime. So this Gaussian-state version of the Fibonacci protocol could overcome distance limitations inherent in quantum cryptography, but at the cost of losing the security gained from the ability to reveal eavesdroppers. In exchange, the high transmission rate of the Gaussian state protocol would allow for sufficient privacy amplification to be implemented in order to bring the information available to Eve back down to a very low level, allowing overall security to be maintained.

\section{Photon-splitting attacks, decoy states, and security}
In QKD, the goal is always to send single photons, but in reality what are sent are attenuated pulses with mean photon number less than one. There will always be some pulses containing more than one photon each. In some standard protocols, such as BB84, all of the photons in the same pulse undergo the same preparation (passing through the same polarizer or same hologram), so they come out with the same polarization or the same OAM. This leads to a serious security problem \cite{huttner,yuen,lutken,brassard}.
Splitting off one photon from each pulse, Eve can measure its state without altering the state of the remaining photons. She sends the remainder of the pulse on to Bob via a channel of lower loss than the original one, in order to mask the fact that she has removed some of the photons. Measures must be taken to safeguard against such photon-number splitting attacks, for example by insertion of decoy states \cite{hwang}.

However, in the Fibonacci protocol, the situation is different. If a multiphoton pulse is sent into the spiral source, there is no reason for the photons that come out to be in the same state: they will be distributed among the different Fibonacci numbers in the same manner as they would if they had been sent in one by one. Siphoning off one photon from a pulse will tell the eavesdropper nothing about the state of the other photons in that pulse. So, the Fibonacci protocol is intrinsically immune to photon-number splitting attacks.

Security can be further enhanced by using a new type of decoy state that is already present automatically in the setup. When the pump photon of OAM $F_n$ produces the down conversion pair with OAM $l_i$ and $l_s$, angular momentum conservation requires $F_n=l_i+l_s$. The bits used to construct the key come from the instances in which this is satisfied by the outgoing angular momenta equalling the two previous Fibonacci numbers: $F_n=F_{n-1}+F_{n-2}$. However, the conservation relation can also be satisfied by values of $l_i$ and $l_s$, {\it neither} of which are Fibonacci (for example: $13=9+4$, in which $13$ is Fibonacci but $4$ and $9$ are not), or in instances where {\it one} of the outgoing values is a Fibonacci number while the other is not (for example, $13=3+10$, in which $13$ and $3$ are Fibonacci, while $10$ is not). Consider the latter case, i.e. $F_n=F_m+l$, where $m\le n-3$ and $l$ is not Fibonacci. If Eve detects the Fibonacci value $F_m$ on its way to Bob, she has no way of knowing that Alice's number is not Fibonacci, so she will send Bob one of the two superpositions ${1\over \sqrt{2}}\left( | F_m\rangle +|F_{m-1}\rangle\right)$ or ${1\over {\sqrt{2}}}\left( |F_m\rangle +|F_{m+1}\rangle \right) $. However, after measurements
are completed, Alice and Bob can exchange information letting each other know for which photons they received Fibonacci or non-Fibonacci values.
If Alice measures non-Fibonacci value $|l\rangle$, then the state arriving in Bob's lab should be ${1\over \sqrt{N}}\sum_n|F_n-l\rangle .$ All but one of the terms in this sum will be non-Fibonacci-valued, with the exception being the term where $F_n-l=F_m$. Suppose Bob compares the state he receives to a test state of the form \begin{eqnarray}|\psi_{test}\rangle &=& {1\over \sqrt{N}}\left\{ |F_{n_0}\rangle - |F_{n_0+1}\rangle + |F_{n_0+2}\rangle
\right. \\ & & \qquad\left. - |F_{n+3}\rangle +\dots \pm  |F_{n_0+N-1}\rangle \right\} ,\nonumber\end{eqnarray} with alternating signs between terms. This state has nonzero inner
product with each individual Fibonacci state, but is orthogonal to all pairwise superpositions of consecutive Fibonacci states. So, when Alice receives a non-Fibonacci number, Bob should expect the inner product between his state and the test state to be ${1\over \sqrt{N}}$ (due to the $F_m$ term in the sum), whereas if Eve has tampered with the state he will instead find the overlap to be equal to zero.
Bob will once again be able to detect Eve's interference through a large (ideally $100\% $) drop in his detection rate for the subset of events where
Alice receives non-Fibonacci values.

Instead of intercept-and-resend or photon-number-splitting attackes, Eve may instead try to clone the state and then send one copy on, keeping the other to measure after intercepting the classical exchange. This, however runs afoul of the no-cloning theorem \cite{wooters}; she can only make perfect clones if the possible states are mutually orthogonal and if she knows what they are. Since the superpositions Eve intercepts are not mutually orthogonal, detectable errors will again be introduced.

A more detailed analysis of security issues in the Fibonacci protocol will appear elsewhere.

\section{Doubling the information capacity}
In the main text, we assumed that only positive orbital angular momentum (OAM) values were used, in order to keep the explanation relatively simple.
The OAM sorters before the detectors in fig. 1 of the main text allow us to divert negative OAM values away from the detectors, keeping only positive signal and idler values. This in turn implies that only positive-OAM pump photons contribute.

However, negative OAM values are also created by the source, at the same rate as the positive values. It is to our advantage to expand the setup to make use of these, rather than letting half of the created photons go to waste. When we do this, we find that the number of bits of key generation per photon can be doubled.

\begin{figure}
\centering
\includegraphics[totalheight=1.8in]{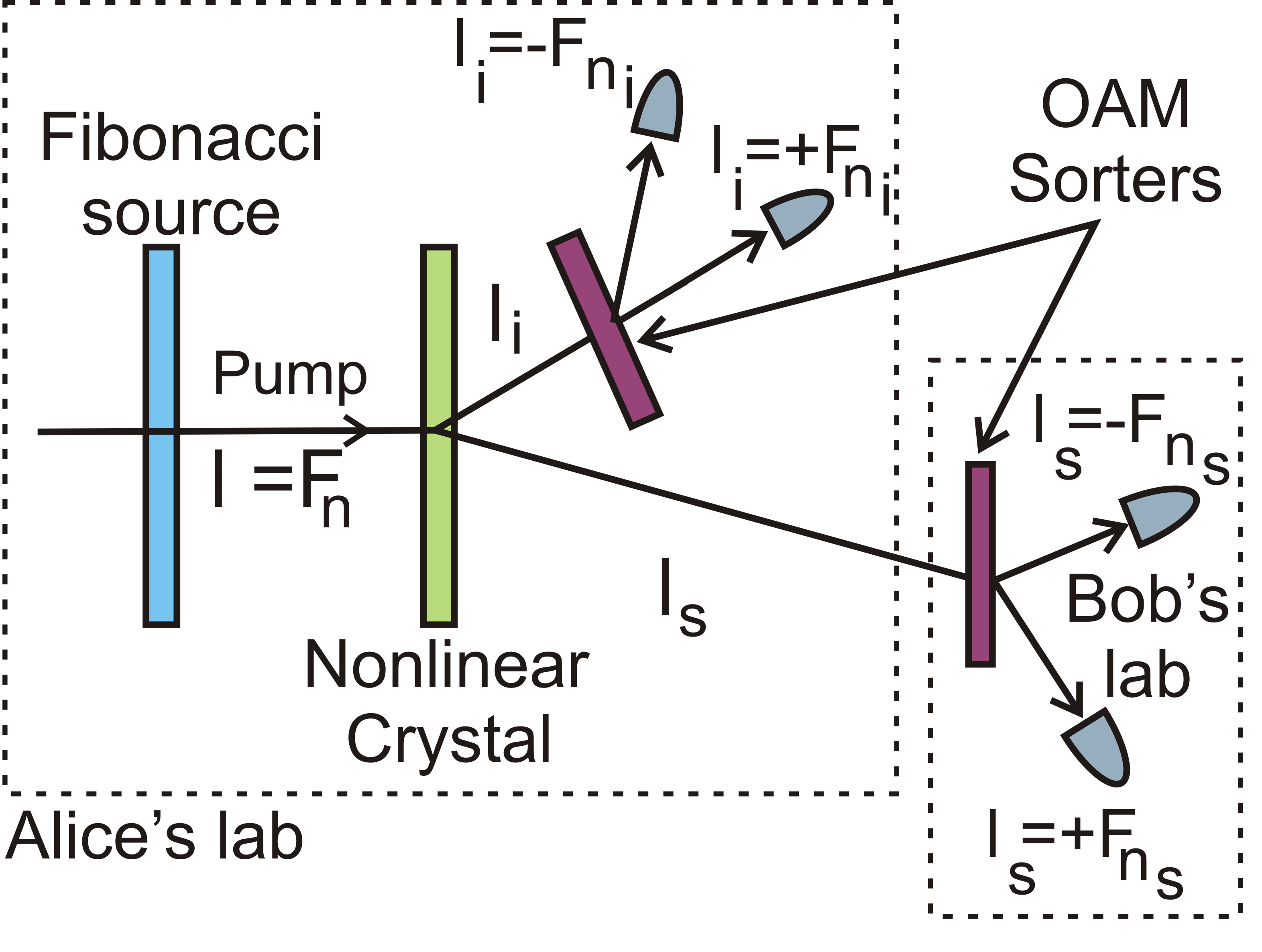}
\caption{\textit{A variation of fig. 1 that makes use of both positive and negative orbital angular momentum values to double key generation capacity. Rather than keeping only positive values, both signs are kept. Alice and Bob must exchange classical information to tell each other the signs of each detection.}}\label{negativelfig}
\end{figure}

A schematic of the expanded setup is shown in fig. \ref{negativelfig}. Alice and Bob count their positive and negative angular momentum values separately, and during the classical exchange let each other know the signs they received. They then only keep the trials on which they received the same signs. (This is in a sense analogous to keeping only the sets of matching bases in the BB84 and E91 protocols.)  Each positive or negative Fibonacci number can then represent a four-digit binary string:

\begin{center}\begin{tabular}{ccc|cc}$l$ & Binary string & \qquad & $l$ & Binary string\\ \hline
$3$ & $0000$ & & $-3$ & $1000$\\
$5$ & $0001$ & & $-5$ & $1001$\\
$8$ & $0010$ & & $-8$ & $1010$\\
$13$ & $0011$ & & $-13$ & $1011$\\
$21$ & $0100$ & & $-21$ & $1100$ \\
$34$ & $0101$ & & $-34$ & $1101$\\
$55$ & $0110$ & & $-55$ & $1110$ \\
$89$ & $0111$ & & $-89$ & $1111$
\end{tabular}\end{center}

So we now have $16$ possible outcomes for the key segment, with each segment capable of encoding $4$ bits of information via a single photon.

{\bf Acknowledgements.} This research was supported by the DARPA InPho program
through US Army Research Office award W911NF-10-1-0404, by the
AFOSR program "Deterministic Aperiodic Structures for On-chip Nanophotonic and
Nanoplasmonic Device Applications" under Award FA9550-10-1-0019, and by NSF Career Award
No. ECCS-0846651. The authors are extremely grateful to
Alexander Roth and Ari B. Roth for their help with the graphics.

\end{document}